\documentclass[aps,prl,twocolumn,showpacs]{revtex4}
\usepackage{epsfig}
\usepackage{graphicx}
\usepackage{color}

\newcommand{\fig}[1]{Fig.~\ref{#1}}

\def\Hz{{\ \text{Hz}}} 
\def\kHz{{\ \text{kHz}}} 

\def\Rb87{^{87}\text{Rb}} 
\def\Na23{^{23}\text{Na}} 
\def\Li6{^{6}\text{Li}} 

\def\bra#1{\mathinner{\langle{#1}|}}
\def\ket#1{\mathinner{|{#1}\rangle}}

\def\Bra#1{\left<#1\right|}
\def\Ket#1{\left|#1\right>}
{\catcode`\|=\active
 \gdef\Braket#1{\left<\mathcode`\|"8000\let|\BraVert {#1}\right>}}
\def\BraVert{\egroup\,\mid@vertical\,\bgroup}

\begin{document}

\DeclareGraphicsExtensions{.eps,.EPS}
\title{Resonant demagnetization of a dipolar BEC in a 3D optical lattice}
\author{A. de Paz$^{1,2}$, A. Chotia$^{2,1}$, E. Mar\'echal$^{2,1}$, P. Pedri$^{1,2}$, L. Vernac$^{1,2}$, O. Gorceix$^{1,2}$ and B. Laburthe-Tolra$^{2,1}$}
\affiliation{$^{1}$\,Universit\'e Paris 13, Sorbonne Paris Cit\'e, Laboratoire de Physique des Lasers, F-93430, Villetaneuse, France\\
$^{2}$\,CNRS, UMR 7538, LPL, F-93430, Villetaneuse, France.}
\begin{abstract}

We study dipolar relaxation of a chromium BEC loaded into a 3D optical lattice. We observe dipolar relaxation resonances when the magnetic energy released during the inelastic collision matches an excitation towards higher energy bands. A spectroscopy of these resonances for two orientations of the magnetic field provides a 3D band spectroscopy of the lattice. The narrowest resonance is registered for the lowest excitation energy. Its line-shape is sensitive to the on-site interaction energy. We use such sensitivity to probe number squeezing in a Mott insulator, and we reveal the production of three-body states with entangled spin and orbital degrees of freedom.

\end{abstract}

\date{\today}
\maketitle

Quantum dipolar gases have attracted much attention in recent years \cite{ReviewBaranov,Baranov2008,ReviewLahaye}. Following seminal studies of Cr Bose Einstein Condensates (BECs) \cite{stuhler2005,lahaye2008,bismut2010,bismut2012},  and the recent production of Er \cite{aikawa2012} and Dy \cite{lu2011,lu2012} quantum gases, it is natural to study dipolar gases confined in optical lattices. Dipolar gases in lattices provide an ideal playground for studying quantum phase transitions in a system with long range interactions. The unique properties of dipole-dipole interactions (DDIs) presents direct similarities with the Heisenberg model of quantum magnetism \cite{Auerbach,Micheli2012,Gorshkov2011,Peter2012}, and lead to novel quantum phases displaying possible long range ordering in lattices \cite{Menotti2007,Lewenstein,HiddenOrder}. The non-linear coupling between spin and orbital degrees of freedom provided by dipolar interactions  \cite{kawaguchi2006,santos2006,Gawryluk2007,sun2007,swislocki2011,PasquiouTubes} is particularly interesting. Although at large magnetic field, this coupling leads to fast dipolar relaxation losses for atoms in excited spin states \cite{PasquiouPRA}, we have recently demonstrated that working at low magnetic field enables the study of multicomponent quantum gases with free magnetization \cite{Demagnetization,Thermodynamics}. Here we extend this research to include dipolar particles trapped in optical lattices, and directly observe discrete, atom-number dependent coupling between spin and orbital degrees of freedom.

We study the magnetization of $^{52}\text{Cr}$ atoms loaded in a deep 3D optical lattice. Under our experimental conditions, we produce a Mott insulator state with a core of two particles per site \cite{Jaksch98, MottGreiner}. We show that on-site dipolar relaxation is inhibited unless the released magnetic energy matches a lattice band excitation, in which case we observe demagnetization resonances. The interplay between the anisotropies of dipolar interaction and of lattice sites leads to a resonance spectrum which depends on the magnetic field orientation. Measuring demagnetization as a function of the magnetic field for two orientations allows for spectroscopy of the 3D lattice band structure. The narrowest resonance is found to be sensitive to the on-site atom number distribution and reveals the number squeezed distribution of the Mott state.  Changing our experimental conditions, we prepare sites containing three atoms. Spin-orbit coupling DDI then produces three-body states that are expected to have entangled spin and orbital degrees of freedom.

\begin{figure}[!t]
   \begin{center}
\includegraphics[width= 6 cm]{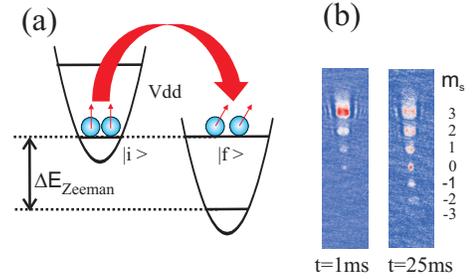}
\end{center}
\caption{\setlength{\baselineskip}{6pt} {\protect\scriptsize
(Color online) 
(a) Sketch: Dipolar relaxation only occurs provided the loss of Zeeman energy $\Delta E_{\rm Zeeman}$ matches a gain in orbital energy (band excitation). (b)  State composition at resonance after 6 ms and after 30 ms of dipolar relaxation (absorption images after Stern-Gerlach separation). 
}}
\label{Figure1}\end{figure}

\begin{figure*}
\centering
\includegraphics[width= 7 in]{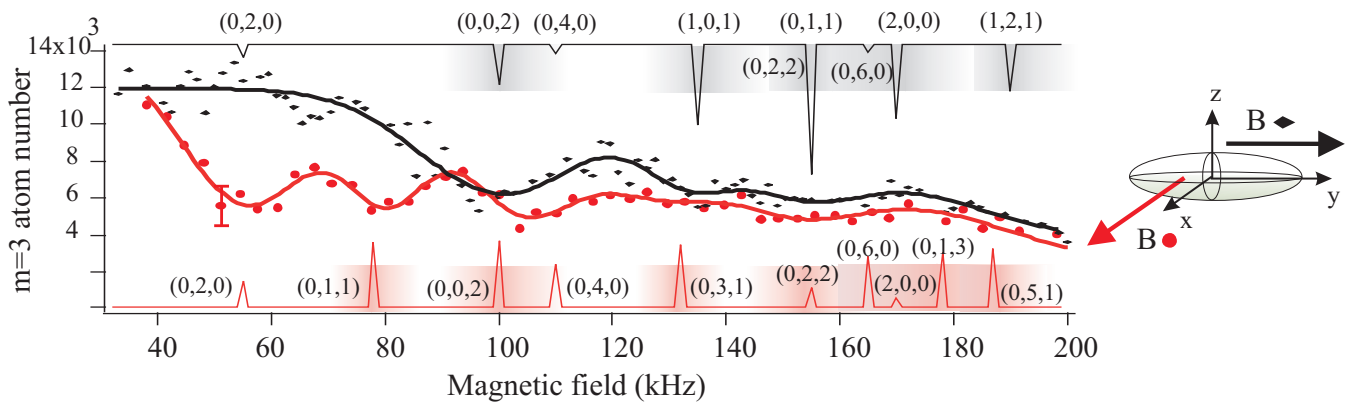}
\caption{\setlength{\baselineskip}{6pt} {\protect\scriptsize
(Color online) Observation of resonances due to dipolar relaxation as a function of the magnetic field amplitude. The spectra (red bullets - after 30 ms of dipolar relaxation time, black diamonds - after 20 ms) are obtained respectively for a magnetic field along $Ox$ and $Oy$, two orthogonal directions in the horizontal plane. The sketch on the right represents the two studied magnetic field orientations referred to a lattice site. The peaks represent the frequency positions and the calculated coupling amplitudes of the resonances leading to excitations of lattice bands labeled by numbers $(n_x,n_y,n_z)$. Shaded areas qualitatively represent energy broadening due to tunneling in excited bands.}}
\label{Figure2}
\end{figure*}

We produce a $^{52}$Cr BEC (with spin $S=3$) with typically $N=1,5 \cdot 10^4$ atoms by forced evaporation in the lowest energy state $\Ket{m_S=-3}$ in a crossed optical dipole trap derived from a fiber laser at 1075 nm \cite{BECParis13} ($m_S$ is the spin projection along the magnetic field). The BEC is loaded adiabatically in a 3D optical lattice, generated using 4 W of a single mode laser at $\lambda$=532 nm.  A 2D horizontal lattice is created by the interference of three beams: two beams are counter propagating, the third intersects the retro-reflected pair at $45^\circ$. The resulting 2D lattice is rectangular with a periodicity of $\lambda / (2\cos \frac{\pi}{8})$ along $Ox$ direction and $\lambda/(2\sin \frac{\pi}{8})$ along  $Oy$ direction. In the vertical direction an independent 1D lattice is created by a vertical retro-reflected beam with horizontal polarization. The $1/e^2$ beams radii at the BEC position are 60(10) $\mu$m. The lattice depth is measured by Kapitza-Dirac diffraction \cite{gould1986} to be 25 $E_R$ (with $E_R= h^2/2 m \lambda^2$ the recoil energy), corresponding to first band excitation frequencies of $\omega_x/ 2\pi=170(10) \kHz$, $\omega_y/2\pi= 55(5) \kHz$ and $\omega_z/2\pi=100(10) \kHz$.

For most of the experiments presented here, we load the optical lattice adiabatically in 40 ms. Although the lattice geometry is complicated by its anisotropy, and we do not expect a fully adiabatic crossing towards the Mott state, this loading time scale allows to remain close to the ground state. This is confirmed by adiabatically ramping down the lattice at the same speed, in which case a BEC with little excitations is recovered. For our experimental parameters, the predicted Mott state is made of a central region with double occupancy surrounded by a single occupancy shell.

Once the lattice is loaded, atoms are adiabatically transferred from $\ket{m_S=-3}\equiv\ket{-3}$ to the highest energy state $\ket{3}$ by means of a 5 ms radio-frequency (rf) sweep (the efficiency of rf sweeps is 85 \%). Once atoms are transferred to $\ket{3}$, dipolar relaxation is allowed. After a given time, the magnetization is again reversed using a second rf sweep (this step is performed only to optimize the detection), the optical lattice is switched off in 1 ms and the populations in the different magnetic states are monitored by a Stern Gerlach analysis (\fig{Figure1} b).

We obtain two different behaviors as a function of the magnetic field amplitude $B$ i.e. whether the Larmor frequency  $\hbar \omega _{\rm Lar}=g_S \mu_B B$ is smaller or larger than the lattice depth $V_0$ ($\mu_B$ is the Bohr magneton and $g_S=2$ is the Land\'e factor). For large magnetic fields, $\hbar\omega_{\rm Lar} > V_0$, we observe a non exponential loss dynamics: the atom number rapidly decreases before reaching a non zero steady value and then no loss are observed for tens of ms. Losses are due to dipolar relaxation between atoms occupying the same site and they stop when only single occupancy sites remain. 

 At low magnetic field, $\hbar\omega_{\rm Lar} \ll V_0$, no losses are observed, as atoms remain trapped after dipolar relaxation. We instead observe a demagnetization dynamics, that reveals a transfer of atoms from $\ket{3}$ to $\ket{2}$ and then to lower energy magnetic states (\fig{Figure1} b). 
For a fixed relaxation time we registered the final population in $\ket{3}$ as a function of the magnetic field amplitude. The spectra presented in \fig{Figure2} show resonance peaks which we interpret as the consequence of total energy conservation: for a pair of atoms initially in the vibrational ground state of one lattice site, spin relaxation only occurs if the released magnetic energy matches a band excitation. We obtain two different spectra for two orthogonal directions of the magnetic field $\mathbf{B}$. This shows that dipolar relaxation couples the initial state to different lattice bands depending on the orientation of $\mathbf{B}$ relative to the geometry of a lattice site.

 To interpret our data, we consider a simple theoretical model with two particles in a purely 3D harmonic anisotropic potential (with frequencies $\omega_x, \omega_y, \omega_z$), describing one lattice site. Dynamics is determined by the dipole-dipole interaction  $ V_{dd}(\mathbf{r})= \frac{d^2}{r^5}[r^2\mathbf{S}_1\cdot\mathbf{S}_2-3(\mathbf{S}_1\cdot\mathbf{r})(\mathbf{S}_2\cdot\mathbf{r})]
$, where $d^2 =\mu_0 \left(g_S \mu_B\right)^2 / 4 \pi$, $\mu_0$ the magnetic constant; $\mathbf{r}=\mathbf{r}_1-\mathbf{r}_2$ is the relative coordinate; $\mathbf{S}_i$ is the spin operator for atom $i$. In presence of an harmonic confinement, the center of mass and the relative motions are decoupled. DDIs affect only the relative motions, therefore only the relative coordinates need to be considered, with eigenstate $\ket{n_x,n_y,n_z}$.  For the magnetic state of the pair, we use the molecular basis $\ket{S_M,m_M}$ where $S_M$ is the total spin and $m_M$ is the spin projection. The energy is given by $E_{{\bf n},m_M}=\sum_{i}{\hbar\omega_i(n_i+\frac{1}{2})}+ m_M g_S \mu_B B$ where $i=x,y,z$.
  The particles are prepared in the initial state $\ket{i}=\ket{0,0,0}\otimes\ket{6,6}$.  

From this initial state, due to DDIs, three spin-relaxation channels are possible with final spin states $\ket{6,5}$, $\ket{6,4}$ and $\ket{4,4}$ \cite{PasquiouPRA}.  The couplings to these three final states are described by the DDIs coupling operators   $V_1=3S^{\frac{3}{2}}d^2 (\hat{x}'+i \hat{y}' )\hat{z}' /\hat{r}^5$, 
$  V_2=\frac{3}{2} \sqrt \frac{6}{11} S d^2 (\hat{x}'+i\hat{y}')^2/\hat{r}^5$ and $  V_3=\frac{3}{2}\sqrt \frac{5}{11} S d^2 (\hat{x}'+i\hat{y}')^2/\hat{r}^5$ where the orientation of the magnetic field $\bf B$ is along the $z'$-axis \cite{PasquiouPRA} ( $x', y'$ and $z'$ are the relative coordinates). The first channel induces a loss of magnetic energy $g_S\mu_B B$, and the two others of $2 g_S\mu_B B$.

\begin{figure}[!t]
\begin{center}
   \includegraphics[width= 8 cm]{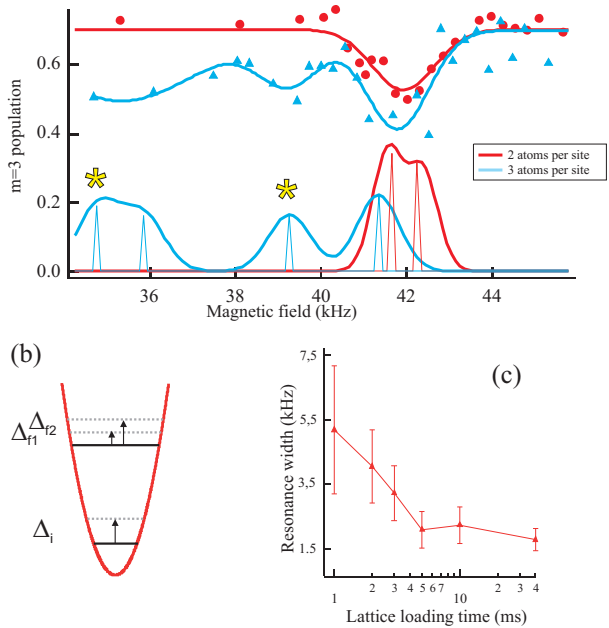}
   \end{center}
\caption{\setlength{\baselineskip}{6pt} {\protect\scriptsize
(Color online) High resolution spectroscopy of the lowest frequency dipolar resonance: sensitivity to site atom number. (a) With two atoms per site (red bullets), the line shape is narrow, while with three atoms per sites (blue triangles), a multi-resonant structure reveals the production of three-body states with entangled spin and orbital degrees of freedom (two peaks with a star). The solid lines (top) are fits, whereas the bottom dashed lines result from calculation (including an empirical broadening mimicking technical noise). Each data point is the average of five measurements. (b) Sketch to interpret the site occupancy sensitivity, arising from an on-site interaction shift depending on the orbital state in the lattice. (c) Resonance width (obtained by fitting the data with a single gaussian) as a function of the lattice loading time.}}
\label{Figure3}
\end{figure}

As sketched in \fig{Figure1} the Zeeman energy is converted into excitation to a higher orbital state $\ket{n_x,n_y,n_z}$. Due to the dependence of $V_{1,2,3}$  on $x',y'$ and $z'$, the selection rules and the coupling strengths to final orbital states strongly depend on the orientation of the magnetic field $\bf B$. This explains the difference between the two registered spectra (see \fig{Figure2}) for $\mathbf{B}$ along $Ox$ and $Oy$ respectively. DDIs couple the initial orbital state with orbital excited states with even total symmetry. Additional specific selections rules are derived from the parity of $V_{1,2,3}$ in the 3 directions $x', y', z'$. In practice, most of the resonances observed in \fig{Figure2} are attributed to channel 2 and 3. The observed resonances corresponds to a rate of the order of 100 $\Hz$ in good agreement with a calculation of the coupling strengths $\Bra{i} V_{1,2,3} \Ket{f}$. The relatives coupling strengths are represented by the  peaks amplitudes in \fig{Figure2} and we also give the their value in supplemental material.

A striking observation in \fig{Figure2} is that resonances widen as the excitation energy increases.  The main source of broadening is the increasing band-width for higher energy lattices bands (represented by the shadows around the peaks in \fig{Figure2}). Due to large lattice periodicity in the $y$ direction, the tunneling rate is small even for the second excited band corresponding to the lowest energy resonance at $\omega_{\rm Lar}=\omega_y=2\pi \cdot 55 \kHz $ but it increases rapidly for excited bands in the other directions (as represented in \fig{Figure2}). Magnetic field and lattice depth fluctuations are broadening the resonances, at the 10 kHz level for data shown in \fig{Figure2}.
Another cause of broadening is cascading dipolar processes corresponding to excitations to higher bands, whose energies are not equally spaced due to anharmonicity (at the 2 kHz level). Cascading is indeed observed in our experiment in which states down to $\Ket{m_S=-2}$ are produced at long time (\fig{Figure1} b). Inhomogeneous broadening of the band structure due the finite size of the lattice beams is of the order of 500 \Hz.

The lowest frequency resonance has a small broadening due to tunneling (100 Hz), allowing  to reveal the role of on-site interactions. This resonance occurs at $\omega_{Lar}=\omega_y$ corresponding to an orbital excitation of 2 units ($2\hbar\omega_y$) and a change in Zeeman energy of $2 g_S\mu_B B$ (the coupling operators are $V_2$ and $V_3$).
In the rest of this letter  we focus on this resonance. 
For these measurements, shown in \fig{Figure3}(a) after 11.5 ms of relaxation, the magnetic field and the laser intensity were actively stabilized to reduce technical broadening. These data were also registered at slightly lower lattice depth than for \fig{Figure2} data (corresponding to $\omega_y=45 (1) \kHz$). 
To explain the observed data, the contact interactions 
$U(\mathbf{r}_1-\mathbf{r}_2)= U_0 \delta(\mathbf{r}_1-\mathbf{r}_2)$  have to be taken into account
($U_0=\sum^{6}_{S_t=0,2,4,6} g_{S_M} P_{S_M} $ , $P_{S_M}=\Ket{S_M}\bra{S_M}$ is the total spin projector, 
$g_{S_M}=4 \pi \hbar^2 a_{S_M}/m$ and $a_{S_M}$ the scattering length for the different molecular channels).
The interactions shift the position of the resonances: for the lowest one we obtain
$\omega_{\rm Lar}=\omega_y+\frac{\Delta_f-\Delta_i}{2} $ (as sketched in \fig{Figure3}(b)).
We calculate the interaction energy shifts of the initial and final states $\Delta_i$ and $\Delta_f$ using first order perturbation theory assuming a harmonic trap.
 
For two particles in the same site, the on-site energy shift of the initial state $\ket{i}$ is given by $\Delta_i=\Delta_0$ where $\Delta_0=\left(m/h\right)^{\frac{3}{2}} (\omega_x \omega_y \omega_z)^{\frac{1}{2}}g_6$. The final state are $\ket{f_1}=\Ket{0,2,0}\otimes\Ket{6,4}$ and  $\ket{f_2}=\Ket{0,2,0}\otimes\Ket{4,4}$. The spin dependent interactions lift the degeneracy and the two final states shifts are
$\Delta_{f1}=\Delta_{0}/2$ and $\Delta_{f2}=(g_4/g_6)\Delta_{0}/2$.

As the initial state is  more confined than the final ones, on-site interactions induce a global red shift of the resonance (about 2.5 kHz) and a splitting of the resonance into two peaks (separated by 0.85 kHz, see  \fig{Figure3} (a)) corresponding to two molecular potentials. We do not resolve the two resonances  for two atoms per site, likely due to broadening associated with residual magnetic field fluctuations and lattice inhomogeneities.

In another set of measurements, we have loaded the lattice faster (in less than 5 ms). 
The line-shape of the resonance is then completely modified, as shown \fig{Figure3}, and presents a multi-resonant structure
that we interpret as due to dipolar relaxation within sites containing three atoms. For atoms in stretched states, the on-site interaction shift scales with the number of pairs. However, it is not the case when the spin degree of freedom is released and the previous result with 2 atoms per site cannot be simply scaled by the number of pairs. For this specific case with three atoms per site we will now show that DDIs couples the initial state to four different fully symmetric states, two of them having entangled spin and orbit. The observed multi-resonant structure originates from these states.

To solve the case of 3 atoms per site, we have used the Jacobi coordinates: the orbital part of the three-body wave function can be described by center of mass coordinate and two relative coordinates (see supplemental material). As in the previous two-body problem, the center of mass motion is decoupled, the total magnetization decreases by two units as the orbital energy increases by $2\hbar \omega_y$. The initial state has maximal magnetization $M=9$ and each particle is in the ground state of the harmonic oscillator. The shift of this initial stretched state is $\Delta_i=3\Delta_0$.  We find four final bosonic states without center of mass excitation, a total magnetization $M=7$ and excitation energy of $2\hbar \omega_y$. The shifts are given by the expressions $\Delta_{f1}=(9/8)\Delta_{0}$,  $\Delta_{f2}=(9/4)\Delta_{0}$,  
$\Delta_{f3}=(a-b)\Delta_{0}$ and $\Delta_{f4}=(a+b)\Delta_{0}$ 
with $a=(150g_4 + 147g_6)/(176 g_6)$ and $b=(5092 g_4^2 + 508 g_4 g_6 + 4201 g_6^2)^{1/2})/(176 g_6)$. It can be noticed that, when $g_4=g_6$, only two resonances are expected.

Two of these four states are of the form $\ket{O}\otimes\ket{S}$, where $\ket{O}$ and $\ket{S}$ are symmetrical states of the
orbital and spin parts respectively. The two others are entangled states of orbital and spin degrees of freedom. Similar states produced by spin changing collisions have also been discussed in the context of Fermi spinor gases \cite{Santos2008}.  

We find a good agreement between our theory and the experiment (see \fig{Figure3}). This shows that we can produce and resolve 3-body states with entangled spin and orbital degrees of freedom in the lattice. This may also enable interesting experiments where correlated tunneling of these exotic states could be studied.

This data is also relevant from the many-body point of view. Indeed, \fig{Figure3} (a) also shows that provided the lattice is loaded slowly, no sites with three atoms are produced. This is the signature that a squeezed atom distribution peculiar to the Mott state has been reached. On the other hand, for a short loading time sites with three atoms are produced. In \fig{Figure3} (c) we show the resulting increase of the resonance width for loading times shorter than 5 ms. This adiabatic timescale is comparable to the trap period, which suggests that the breakdown of adiabaticity for fast loading arises from a time scale that is so short that the cloud does not have time to swell to reach the ground state of the
system.

In conclusion, we have observed anisotropic dipolar resonances that couple the spin and the orbital degrees of freedom of bosons in a 3D lattice. We have focused on the narrowest resonance which can be used to engineer few-body entangled states and to probe the atom number distribution in the Mott state. Dipolar resonances produce correlated states whose tunneling properties would be interesting to study. An interesting prospect would also be to reach the coherent regime. For example, starting from atoms in $\Ket{-2}$ one might expect Rabi oscillations coupling states of different magnetization and different orbital states. Finally, we also stress that away from dipolar resonances, the magnetization is fixed, and the gas is stable. The system is then a good candidate to study lattice magnetism at constant magnetization, with interesting connections to quantum magnetism \cite{Barnett2006}. The magnetic stability away from dipolar relaxation resonances displays similarities with the stability of quantum dots discussed in the context of quantum information computing \cite{paillard2001}.
\vspace{0.5cm}

Acknowledgements: LPL is Unit\'e Mixte (UMR 7538) of CNRS and of Universit\'e Paris 13. We acknowledge financial support from Conseil R\'{e}%
gional d'Ile-de-France, Minist\`{e}re de l'Enseignement Sup\'{e}rieur et de la Recherche (CPER) and IFRAF. A. C. thanks CNRS for financial support. We thank J.V.~Porto and M.~Gajda for their suggestions and their comments during the preparation of the manuscript.



\newpage
\null
\centerline{\Large \textbf{Supplemental material}}

\vskip 5 mm

\paragraph{Transition matrix elements.}

Here we present the result of the calculation of the matrix elements related to the dipolar relaxation resonances involving two atoms per site. The site frequencies are  $\omega_x/ 2\pi=170 \kHz$, $\omega_y/2\pi= 55 \kHz$ and $\omega_z/2\pi=100 \kHz$. Channel 1 corresponds to the final spin state  $\Ket{6,5}_{\rm mol}=\frac{1}{\sqrt{2}}(\Ket{3,2}_{\rm ato}+\Ket{2,3}_{\rm ato} )$.  Channel 2 and 3 that correspond to final spin states $\Ket{6,4}_{\rm mol}$ and $\Ket{4,4}_{\rm mol}$ respectively are almost degenerated. For data presented in Fig.2 , the splitting due to contact interaction is not resolved. Therefore we present in the Table the coupling to the atomic final state $\Ket{2,2}_{\rm ato}=\sqrt{\frac{6}{11}}\Ket{6,4}_{\rm mol}-\sqrt{\frac{5}{11}}\Ket{4,4}_{\rm mol}$ instead of the coupling to the molecular spin states. 
The final state $\Ket{n_x,n_y,n_y}$ corresponds to the excitation of the relative motion along the three different axis. We give the frequencies of the resonnance and the couplings $\Bra{f} V_{dd} \Ket{i}$ where $\ket{i}=\ket{0,0,0}\otimes\ket{6,6}$ is the initial state. 
\begin{center}
\footnotesize
\begin{tabular}{c|c|c|c}
\hline
  \multicolumn{4}{c}{\normalsize channel 1}\\
\hline
 Final & Resonance&Coupling &Coupling \\
 state& (kHz)&B$\parallel$Ox (Hz)&B$\parallel$Oy (Hz)\\
\hline
 $\ket{0, 2, 0}$& 110& 0 &0\\
 $\ket{0, 1, 1}$& 155& 0&412\\
 $\ket{0, 0, 2}$& 200& 0&0\\
 \hline
\multicolumn{4}{c}{\normalsize channel 2 \normalsize \& 3}\\
\hline
 Final & Resonance&Coupling &Coupling \\
 state& (kHz)&B$\parallel$Ox (Hz)&B$\parallel$Oy (Hz)\\
\hline
 $\ket{0, 2, 0}$& 55& 107 &47\\
 $\ket{0, 1, 1}$& 77.5& 238&0\\
 $\ket{0, 0, 2}$& 100& 230&152\\
 $\ket{0, 4, 0}$& 110& 164&35\\
 $\ket{1, 1, 0}$& 112& 0&0\\
 $\ket{0, 3, 1}$& 132.5& 227&0\\
 $\ket{1, 0, 1}$& 135& 0&305\\
 $\ket{0, 2, 2}$& 155& 62&84\\
 $\ket{0, 6, 0}$& 165& 190&28\\
 $\ket{1, 3, 0}$& 167.5& 0&0\\
 $\ket{2, 0, 0}$& 170&38 &278\\
 $\ket{0, 1, 3}$& 177.5&208 &0\\
 $\ket{0, 5, 1}$& 187.5&223 &0\\
  $\ket{1, 2, 1}$& 190& 0&176\\
  \hline
\end{tabular}
\end{center}

\paragraph{Calculation details for coupled three body states.}

We present here the calculation of the dipolar relaxation resonance spectrum (see Fig. 3) when 3  atoms are loaded in the same site. Initially, 3 atoms occupy the ground state of an anisotropic harmonic site with frequencies $\omega_x,\omega_y$ and $\omega_z$. DDIs induce magnetization changing collisions with excitations to higher orbital states. We consider in the problem only orbital excitations along the $Oy$ direction. Indeed, under our experimental conditions, $\omega_y$ is the smallest frequency and the registered lowest frequency resonance corresponds to excitation of the $2\hbar\omega_y$ (without central of mass excitation) that comes with a decrease of the total magnetization by two units. We describe the three body states by means of the states $\ket{n_1,n_2,n_3;m_1,m_2,m_3}$ where $n_i$ is the $Oy$ harmonic oscillator quantum number and $m_i$ the spin projection for particles $i=1,2,3$. 

The orbital energy is $E=\hbar(n_1+n_2+n_3)\omega_y+E_0$ and total magnetization $M=m_1+m_2+m_3$. The initial state is $\ket{i}=\ket{0,0,0;3,3,3}$ with energy $E_0$.

The orbital final states with $\Delta E = 2 \hbar \omega_y$ are $\ket{2,0,0}$, $\ket{0,2,0}$, $\ket{0,0,2}$, $\ket{1,1,0}$, $\ket{1,0,1}$ and $\ket{0,1,1}$. With these states, we can construct 3-body states without center of mass excitations. For this purpose, we use Jacobi coordinates $R=\frac{y_1+y_2+y_3}{3}$, $r=(y_1-y_2)$, $\rho=\frac{y_1+y_2}{2}-y_3$ and we construct three operators $\hat{a}_R=\frac{\hat{a}_1+\hat{a}_2+\hat{a}_3}{\sqrt{3}}$, $\hat{a}_r=\frac{\hat{a}_1-\hat{a}_2}{\sqrt{2}}$ and $\hat{a}_\rho=\sqrt{\frac{2}{3}}\left(\frac{\hat{a}_1+\hat{a}_2}{2}-\hat{a}_3\right)$ where $\hat{a}_i=\sqrt{\frac{m\omega_y}{2\hbar}}\left(\hat{y}_i+\frac{i}{m\omega_y}\hat{p}_{y_i}\right)$. The first operator acts on the center of mass coordinates, and the two others solely on the relative motion. Using these operators,  the hamiltonian now reads $H=\hbar \omega_y(\hat{a}^\dagger_R\hat{a}^\dagger_R+\hat{a}^\dagger_r\hat{a}^\dagger_r+\hat{a}^\dagger_\rho\hat{a}^\dagger_\rho)+E_0$. From the orbital ground state $\ket{0,0,0}$ we can apply $\frac{(\hat{a}^\dagger_r)^2}{\sqrt{2}}$, $\frac{(\hat{a}^\dagger_\rho)^2}{\sqrt{2}}$ and $\hat{a}^\dagger_r\hat{a}^\dagger_\rho$ to 
generate three orthogonal states without center of mass excitations (these three states are a linear combination of the six final orbital states listed just before).

The initial spin state is $\ket{3,3,3}$ with magnetization $M=9$. The possible final spin states having magnetization $7$ are $\ket{3,3,1}$, $\ket{3,1,3}$, $\ket{1,3,3}$, $\ket{3,3,2}$, $\ket{3,2,3}$ and $\ket{2,3,3}$. Combining the three orthogonal orbital states, we thus obtain $18$ orthogonal 3-body states with $E = 2 \hbar \omega_y+E_0$, no center of mass excitations and total magnetization $M=7$.
These states describe distinguishable particles. We symmetrize these states in order to satisfy bosonic statistics and we finally obtain 4 orthogonal bosonic states without center of mass excitation.

\paragraph{Effect of contact interactions.}

These states are degenerated without interaction. Spin dependent contact interaction fully remove the degeneracy. For three particles, it reads $U(r_1,r_2,r_3) = U_{12}\delta(\mathbf{r}_1-\mathbf{r}_2)+U_{23}\delta(\mathbf{r}_2-\mathbf{r}_3)+U_{13}\delta(\mathbf{r}_1-\mathbf{r}_3)$ where $U_{ij}=\sum^{6}_{S_{ij}=0,2,4,6} g_{S_{ij}} P_{S_{ij}} $ , $P_{S_{ij}}=\Ket{S_{ij}}\bra{S_{ij}}$ is the total spin projector for the particles $i$ and $j$. $g_{S_M}=4 \pi \hbar^2 a_{S_M}/m$ and $a_{S_M}$ the scattering length for the different molecular channels.  For chromium $a_6=102.5$ $a_0$ and $a_4=64$ $a_0$ [24] where $a_0$ is the Bohr radius. Within first order perturbation theory, we diagonalize analytically the 4x4 Hamiltonian, obtaining the energy shifts $\Delta_{f_n}$ reported in the paper and the final orthogonal states  $\Ket{f_n}$. Among the four states two of them are product state of orbit and spin, the two others are entangled states of orbit and spin. With the experimental parameters corresponding to Fig. 3, we obtain  $(\Delta_{f_1},\Delta_{f_2},\Delta_{f_3},\Delta_{f_4})=(20.7, 16.8, 10.4, 8.4)\: {\rm kHz}$

DDIs couple the initial state to the four final states. 
The matrix elements of the dipole-dipole interaction are $\langle f_n |3V_{dd}|i\rangle=\alpha_n\mathcal{V}$.
We obtain  $\mathcal{V}=185 \;{\rm Hz}$ and $(\alpha_1,\alpha_2,\alpha_3,\alpha_4)=(0.51,0.43 , 0.44 , 0.59)$.

\end{document}